\documentclass[]{interact}

\RequirePackage[T1]{fontenc}
\usepackage{array}
\usepackage{tabularx}
\usepackage{multirow}

\usepackage{subcaption}
\RequirePackage{graphicx}
\RequirePackage{mathptmx}      
\RequirePackage{flushend}
\RequirePackage[numbers,sort&compress]{natbib}

\usepackage[english]{babel}
\usepackage{blindtext} 

\usepackage[dvipsnames]{xcolor}

\def\tsc#1{\csdef{#1}{\textsc{\lowercase{#1}}\xspace}}
\tsc{WGM}
\tsc{QE}
\tsc{EP}
\tsc{PMS}
\tsc{BEC}
\tsc{DE}

\begin{document}


\title {Deep Learning Methods for Screening Pulmonary Tuberculosis Using Chest X-rays}


\author{
\name{Chirath Dasanayaka \textsuperscript{a}\thanks{ Email:chirath.dasanayaka@eng.pdn.ac.lk} and Maheshi Buddhinee Dissanayake  \textsuperscript{a}\thanks{CONTACT Maheshi Buddhinee Dissanayake, orcid=0000-0001-5209-5441, principle investigator and the corresponding author at Email: maheshid@eng.pdn.ac.lk}} 
\affil{\textsuperscript{a}Department of Electrical and Electronic Engineering, Faculty of Engineering, University of Peradeniya, Sri Lanka}
}
\maketitle
\date{}




%

\begin{abstract}
Tuberculosis (TB) is a contagious bacterial airborne disease, and is one of the top 10 causes of death worldwide. According to the World Health Organization (WHO), around 1.8 billion people are infected with TB and 1.6 million deaths were reported in 2018. More importantly, 95\% of cases and deaths were from developing countries. Yet, TB is a completely curable disease through early diagnosis. 
To achieve this goal one of the key requirements is efficient utilization of existing diagnostic technologies, among which chest X-ray is the first line of diagnostic tool used for screening for active TB.
{The presented deep learning pipeline consists of three different state of the art deep learning architectures, to generate,} segment and classify lung x-rays. Apart from this image preprocessing, image augmentation, genetic algorithm based hyper parameter tuning and model ensembling were used to to improve the diagnostic process.
We were able to achieve classification accuracy of 97.1\% (Youden’s index-0.941, sensitivity of 97.9\% and specificity of 96.2\%) which is a considerable improvement compared to the existing work in the literature.
 In our work, we present an highly accurate, automated TB screening system using chest X-rays, which would be helpful especially for low income countries with low access to qualified medical professionals.
\end{abstract}



\begin{keywords}
Machine Learning, Automated  Tuberculosis  Screening,  Chest/Lung x-ray analysis, { Deep Convolutional Generative Adversarial  Networks}, Segmentation and Classification  
\end{keywords}

\section{Introduction}

Tuberculosis (TB) is an infectious disease affecting about one-quarter of the world’s population \cite{r1} according  to the WHO, and it is commonly caused by a bacterium known as Mycobacterium tuberculosis. TB is most prevalent in sub-Saharan Africa, Southeast Asian region and Latin America due to poverty,  malnutrition and hunger \cite{r2}. These developing countries have twelve times lesser radiologists than developed countries \cite{r1}, so TB patient often remains undetected while continuing to spread the bacterium further, through the air by coughing, sneezing and spitting. By automating chest X-ray
(CXR) screening process we can tackle this issue. Even though there exists several deep learning based TB detection models in the literature, the accuracy of most of the models are low in average around 80\% \cite{r3,r4,r5,r6}. CAD4TB is a
commercially available product for automatic TB detection and classification \cite{r6}. This software is based on deep learning with shape detection and textural abnormality detection \cite{r3}. The main drawback of this software is low accuracy of 84\% \cite{r7}. In \cite{r3} authors proposed a method based on local binary pattern (LBP) and Laplacian of Gaussian (LoP) with rib suppression to detect TB nodules in CXR images and they were able to achieve an accuracy of 82.8\% on their test set. In \cite{r4}  the authors proposed a method to locate focal opacities of TB in CXR using image processing, boundary tracing and edge detection techniques and they have achieved 85\% of accuracy in their model. In another work \cite{r5} a method based on Bayesian classification was proposed to detect TB  cavities in CXR and they were able to achieve 82.35\% of accuracy on their test set. Almost every model available in the literature has followed the same traditional approaches \cite{r3,r4,r5,r6} and was able to achieve considerable accuracy. Sensitivity is a crucial parameter in medical diagnosing. It is very important, not to misdiagnosis a TB positive patient as a TB negative patient as it results in further spreading of the disease. Besides, having high specificity is also an important factor when screening a large population, as it helps to reduce the workload of medical professionals. Our main objective is to increase the sensitivity and specificity of TB detection process which use only frontal CXR. It is highly appropriate to have an inexpensive, reliable model that can be easily accessible even from remote areas in order to limit the prevalence of TB disease especially in countries with the lowest per capita GDP.\par In our work, we present a model that can detect TB with high sensitivity (97.9\%) and specificity (96.2\%). The pipeline of our work is as follows; {First the TB positive chest x-ray images were synthesized using deep generative adversarial network. From the generated images a set of best images were selected depend on the score of both subjective and objective quality assessment metrics. Peak signal to noise ratio was selected as the objective quality assessment metric, and the subjective evaluation was done by the radiologists.} After that the images were preprocessed to suppress unwanted distortions and enhance important features. Next the UNET convolutional neural network \cite{r8} was used to segment the lung fields. By segmenting the lung fields we further suppressed the unwanted features from the image. Currently deep convolutional neural networks (DCNN) are the state of the art for image classification, but the quality of the DCNN is highly depended on its hyper-parameters. In our work the optimization of the hyper parameters was achieved by using genetic algorithm \cite{r9}. Finally, the ensemble of VGG16 \cite{r10} and InceptionV3 \cite{r11} pre trained models were used to perform the binary classification. The performance of the model was evaluated using sensitivity, specificity and Youden’s index. To get an insight about models behavior and performance we used class activation map (CAM) \cite{r12}. It checks whether the model focus on right features. The final results show that the data volume, image preprocessing and augmentation techniques, lung segmentation, hyper parameter tuning and model ensembling have highly impacted on the final performance of the model. Each of the technique that we used will be discussed thoroughly in relevant sections.  

\section{Data sets}
The two datasets of chest radiography that were used in this work, are available in US National Library of Medicine, and they were acquired from the Department of Health and Human Service, Montgomery Country, Maryland, USA, and Shenzhen No 3 Peoples Hospital in China \cite{r13}. Both datasets were de-identified by the data providers and were exempted from IRB review at their respective institutions. At, NIH the dataset use and public release were exempted from IRB review by the NIH office of Human Resource Research Projection Programs (No. 5357).

\subsection{Shenzhen China Dataset}
This publically available dataset of frontal chest X-rays were collected by Shenzhen No-3 hospital in Shenzhen, Guangdong, China and consists of 662 frontal CXR images, of which 335 images are TB positive and 327 images are TB negative. Exploratory data analysis shows that it is a well-balanced dataset which covers almost every age group (from 1 months old to 75 years old) and distributed almost equal in gender as well (60\% male). Every image is in the format of png and resolution varies from 998x1130 to 3001x3001. Lung masks for this dataset were provided by National technical university of Ukraine \cite{r13}.

\subsection{Montgomery Country Dataset}
This publically available dataset of frontal chest X-rays were provided by Department of Health and Human Services of Montgomery Country, MD, USA and it consists of 138 frontal CXR images, of which 58 images are TB positive and 80 images are TB negative. Every image in the dataset are in png image format and the resolution of the images is either 4020x4892 or 4892x4020. This dataset itself includes lung masks \cite{r13}.\\
{Noted that the X-rays in the datasets with TB negative  are used purely for reference purpose in this study, and they are of healthy individuals.} 

{\subsection{X-Ray Image Synthesis}
It is a well-known fact that the deep learning models are data driven algorithms and it is proven that the accuracy of the model increase as the volume of data and the heterogeneity of data increases. One of the significant challenge in the domain of medical imaging is the scarcity of the reliable and class balance dataset due to various reasons like ethical issues \cite{r14} To overcome this, deep convolution generative adversarial network (DC GAN) was \cite{r15} used. Figure 1 shows the high level architecture of the DC GAN used and figure 2 shows the progression of the generated images against each iteration.} 

\begin{figure}[]
  \centering
	  \includegraphics[width=1.0\linewidth]{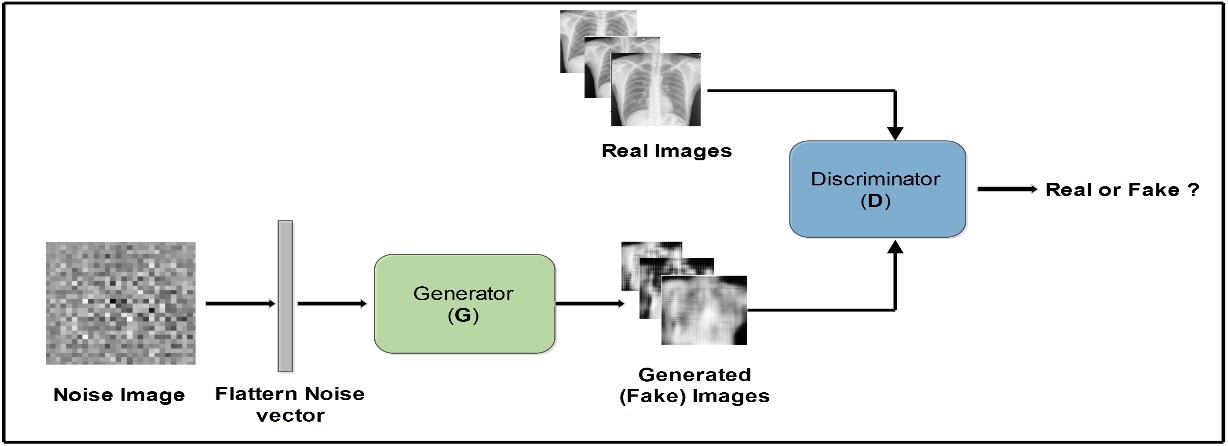}
  \caption{High-level Deep Convolutional Generative Adversarial Network Architecture}
  \label{GenerativeNetwork}
\end{figure}

\begin{figure}[ht] 
  \begin{subfigure}[b]{0.5\linewidth}
    \centering
    \includegraphics[width=0.75\linewidth]{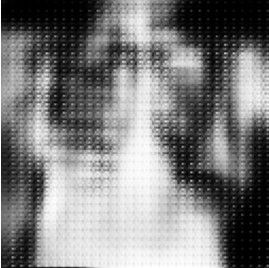} 
    \caption{Generated image after 10 epochs} 
    \label{fig7:a} 
    \vspace{4ex}
  \end{subfigure}
  \begin{subfigure}[b]{0.5\linewidth}
    \centering
    \includegraphics[width=0.75\linewidth]{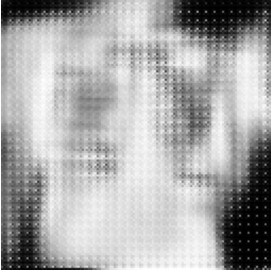} 
    \caption{Generated image after 100 epochs} 
    \label{fig7:b} 
    \vspace{4ex}
  \end{subfigure} 
  \begin{subfigure}[b]{0.5\linewidth}
    \centering
    \includegraphics[width=0.75\linewidth]{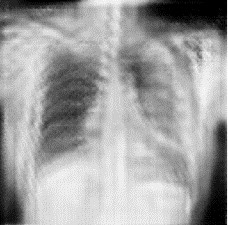} 
    \caption{Generated image after 1000 epochs} 
    \label{fig7:c} 
  \end{subfigure}
  \begin{subfigure}[b]{0.5\linewidth}
    \centering
    \includegraphics[width=0.75\linewidth]{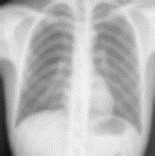} 
    \caption{Generated image after 6000 epochs} 
    \label{fig7:d} 
  \end{subfigure} 
  \caption{Progression of the generated images with epoch}
  \label{fig7} 
\end{figure}

{The resolution of the output generative images were set to 128x128. Because as with the increase of the resolution it becomes easy for discriminator to decide whether the generated image is real or fake. This cause mode collapsing more often. By using the DC GAN, 1000 TB positive frontal x-rays were generated and selected according to the quality evaluation metrics (Objective- PSNR, Subjective- Medical expert opinion) and another 970 of healthy x-rays were collected from MIMIC database \cite{r16}. Due to this approach, the total frontal chest x-ray images considered for the study was expanded up to 2770 images. Out of which 1393 of them were TB positive and the rest of 1377 were TB negative. Further, in the generated database 1770 images out of 2770 were acquired directly through x rays (real images) and the rest were generated using the DC GAN. 
Then from the combined database 80\% (2216) images were randomly selected for training and 10\% (277) of images were selected as validation set. We choose another 10\% (277) images randomly out of the real image set acquired through references \cite{r13,r16} as our testing set. By selecting real images as testing dataset we mimic a more practical scenario for testing our model presented in this paper.} 

\begin{table}
\centering
\caption{Data sets summary}
\begin{tabular}{|l|l|l|l|l|}
\hline
  Data set & TB positive & TB Negative & Total\\ \hline
MIMIC \cite{r16}      & 0                                                     & 970       & 970     \\ \hline
Montgomery Country \cite{r13} & 58                                                     & 80       & 138       \\ \hline
Shenzhen China \cite{r13}   & 335                                                     & 327       & 662      \\ \hline
Generated   & 1000                                                     & 0       & 1000      \\ \hline
Database   & 1393                                                     & 1377       & 2770      \\ \hline
\end{tabular}
\end{table}


\section{Methodology}

Our approach for detecting TB is discussed in this section. Our proposed method consist of three main sub models. The first sub model was to synthesize TB positive frontal chest x-rays, the second model was to obtain highly accurate lung segmentation and the last sub model was trained for the classification.

{In the implementations, the proposed  CNN models will follow 3 key stages, training, validation and testing.  We divide the entire training set into 80\%, 10\%, 10\% ratio for the above 3 categories respectively. First the designed model is trained for the given task using the training set images. Then the validation set was used to check the unbiased evaluation of a model fit on the training dataset. i.e. it helped us to get an early estimate of the performance of the model while giving clear indication on whether the model has overfitted or not. Finally, the test dataset is used to assess the full performance of the proposed model.}

According to the exploratory data analysis, it can be seen that there are variations in image resolution, size, contrast, and zoom on lungs. The resolution of the images were converted to 128x128 before feeding into the model. However contrast variation could make it hard for the model to learn right features properly. To overcome this bottleneck, histogram equalization \cite{r17} can be used to enhance the contrast of the images but it tends to increase the noise level of the images as well. {Since the DC GAN generated images may already associate with certain amount of noise, histogram equalization could make it worse. Hence, in the proposed model, CLAHE (contrast limited adaptive histogram equalization) \cite{r18} technique was adopted to enhance the contrast of the images while limiting the amplification of noise.} Figure 3 presents a comparison between the two techniques.

\begin{figure}[]
  \centering
\begin{subfigure}{.5\textwidth}
  \center  
  \includegraphics[width=.6\linewidth]{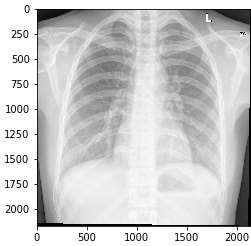}  
  \caption{ Randomly selected row input image.}
  \label{fig:sub-first2}
\end{subfigure}\hfill
\begin{subfigure}{.5\textwidth}
  \centering
    \includegraphics[width=.6\linewidth]{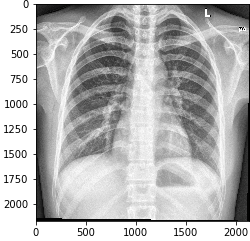}  
  \caption{Histogram equalization of the input image.}
  \label{fig:sub-second2}
\end{subfigure}
\begin{subfigure}{.5\textwidth}
  \centering
  \includegraphics[width=.6\linewidth]{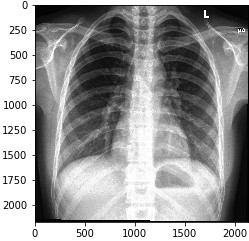}  
  \caption{CLAHE of the input image}
  \label{ fig:sub-three2}
\end{subfigure}
\caption{Comparison of histogram equalization methods.}
\label{fig:fig}
\end{figure}



After the contrast enhancement process noise removal process was applied by using median filtering with a window size of 3x3. Next the images were resized to match the input resolution (128x128x3) of the model. Before feeding the resulting images into DCNN, normalization process was carried out. Images were normalized based on the mean and standard deviation of the images in ImageNet \cite{r19} dataset. This step is quite helpful for training the model in terms of accuracy and learnability.

\subsection{Lung Segmentation}

Lung segmentation is the next preliminary step of our image classification task. Excluding the area that are not pertinent to lung fields, would  help the model to focus more on lung fields and learn correct features while training. The system accuracy and efficiency highly depend on the accurate segmentation result. An accurate lung segmentation is a quite challenging task due to the presence of homogeneous anatomical structure around the lung region. A lot of classical  segmentation techniques are already available in literature such as threshold-based \cite{r20,r21}, region growing \cite{r22}, graph cut algorithm \cite{r23}. However, the current state of the art algorithm for semantic segmentation is UNET convolutional neural network (CNN). For our design we have used UNET architecture since it is known to perform well with Bio medical data. 

\subsubsection{UNET Architecture}

Our proposed network for lung segmentation is based on UNET architecture, which consists of  two parts, namely down sampling followed by up sampling as shown in the figure 4. The down sampling path consist of two 3x3 fully connected convolutions followed by rectified linear unit (ReLU) as the activation function and a 2x2 max pooling operation with stride 2 and padding 1.  
The up sampling path is identical to the down sampling path. To reduce the potential of overfitting due  to the scarcity of the CXR images, heavy data augmentation techniques such as rotation, translation, shearing, horizontal flipping. were applied by the model on the input.

\begin{figure}[]
  \centering
	  \includegraphics[width=1.0\linewidth]{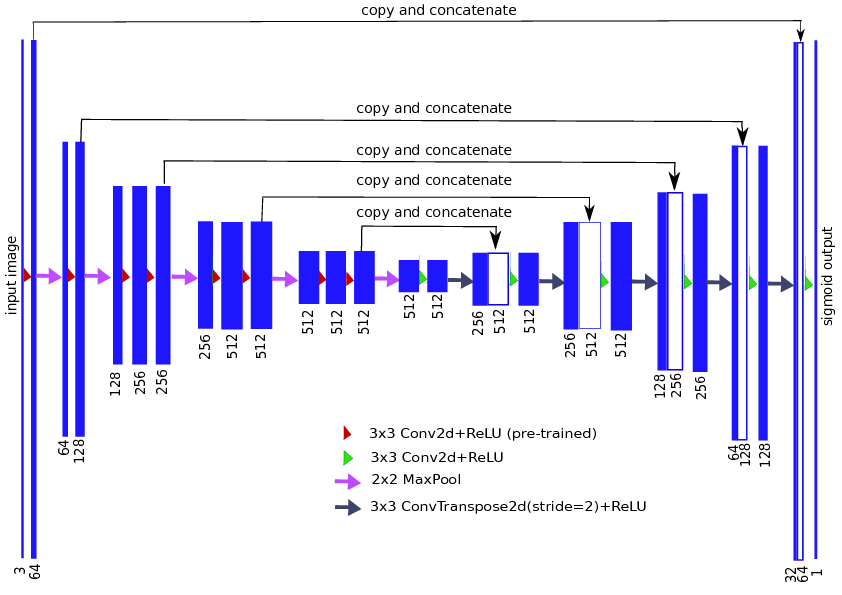} 
  \caption{UNET Architecture used in the model}
  \label{UNET}
\end{figure}

%

\subsubsection{Loss Function}

In image semantic segmentation, the main task of an effective loss function is to improve the discriminative capability of the model. Dice loss is the most popular loss function for medical segmentation and it was chosen for our model as the loss function. Dice loss uses dice similarity coefficient (DSC) \cite{r24} to generate training loss. DSC is a statistical tool used for comparing two sets and can be written as,

\begin{equation}
    DSC(GS,SEG) = \frac{2|GS \cap SEG| }{|GS|+|SEG|} \mbox{\ } \mbox{\ }
\end{equation}

where GS is the gold standard segmentation of the region, SEG represents the corresponding automatic segmentation, and denotes the intersection of the two regions. The metric (DSC) lies within the range of 0 to 1, where 0 represents the worst performance while the 1 represents the best performance. Further, the dice loss is defined as,

\begin{equation}
    Dice Loss = 1- DSC \mbox{\ } \mbox{\ }
\end{equation}

\subsection{DCNN architectures and Transfer learning}
Next we use DCNNs as a classification method to identify TB infected CXRs. First we have trained our data on two pre-trained DCNNs separately. They are VGG16 and InceptionV3. Then the results of the two networks alone and ensemble of the two networks were evaluated. For both pretrained networks we have added two dense layers and an output layer at the top of the network. Then, only the last 2 dense layers were trained while freezing the rest of the layers. This was carried out to limit the training parameters. The input to each of these networks was CXR images, that have been preprocessed and lung field segmented. The overfitting issue was taken care of by increasing the dataset, data augmentation and by limiting training parameters of the DCNN network. {The data augmentations were carried out by using off the shelf image augmentation library “Albumentation” \cite{25}. It has been noticed that the augmentation methods such as grid distortion and elastic transformation along with other blurring and cutout augmentation techniques help greatly to increase the accuracy of the model by a considerable margin.} 
However DCNN model optimization is one of the toughest and crucial factors in deep learn and it can highly impact on the model performance. We choose five hyper parameters for tuning the network which are learning rate, decay factor, momentum, batch size and dropout rate. Then we use a genetic algorithm to figure out the best parameters. Finally, models were trained using the best hyper parameters, and ensembling was performed by taking different weighted averages of the probability scores generated by the each classifiers \cite{r26}.

\subsubsection{DCNN architectures and Transfer learning}

The genetic algorithm is an optimization method which is inspired by the biological evolution. The genetic algorithm repeatedly modifies a population of individual solutions. At each step, the genetic algorithm selects best individuals from the current population to be parents and uses them to produce the children for the next generation. The population evolves toward an optimal solution over successive generations. The three main genetic operators of genetic algorithm are,

    \begin{enumerate}
       \item Selection.
       \item Crossover.
       \item Mutation.            
    \end{enumerate}

{In \textit{selection} we choose best individuals from the population depending on their fitness score. These individuals are the parents of the next generation. In \textit{crossover} we combine selected best parents to form the next generation. \textit{Mutation} applies random changes to the hyper parameters of the parent networks. In our case we choose \textit{validation accuracy} as our fitness function over \textit{testing accuracy} as it is computationally efficient and also gives early estimation about the behaviour of the model. We generate 20 DCNN’s with randomly initialized hyper parameters. After the 30th epoch the best 5 networks were chosen based on their fitness scores which is the validation accuracy at the 30th iteration. Other 15 networks were killed. Using the selected 5 networks another 15 networks were generated. We continue this process over 30 generations. Best parent with highest fitness value after generation 1 and 30 for the VGG16 network is tabulated in Table 2. Parent 1 denotes the best network hyper parameters after each generation.}

\vspace{3mm}

{\textbf{Table 2} Hyper Parameter Optimization}

\begin{tabularx}{0.6\textwidth} { | >{\raggedright\arraybackslash}X | >{\centering\arraybackslash}X | >{\raggedleft\arraybackslash}X | }
   \hline
   Hyper Parameters & \small Hyper parameters of parent 01 after generation 01 \par & \small Hyper parameters of parent 01 after generation 30 \par \\
   \hline
   Learning Rate  & 0.0018 & 0.0007\\
   \hline
   Decay Factor  & 1.3e-5 & 1.1e-5\\
   \hline
   Dropout rate  & 0.7 & 0.3\\
   \hline
   Batch size  & 8 & 16\\
   \hline
   Momentum  & 0.6 & 0.9\\
   \hline
\end{tabularx}

\subsection{Evaluation Metrics}

We randomly choose 277 images (10\%) from \textit{Shenzhen China}, \textit{Montgomery Country} \cite{r13} and MIMIC \cite{r16} datasets to form our testing data set and it consist of 130 tuberculosis negative patients and 147 tuberculosis positive patients. To assess the overall performance following evaluation metrics were used. They are accuracy, sensitivity, specificity, and Youden’s index (YI). Accuracy is a performance evaluation metric that was widely used to asses classification models. But accuracy is sensitive to the class imbalances. Another drawback of accuracy is that two classifiers can yield the same accuracy but perform differently with respect to the types of correct and incorrect classification they provide. However, we have used accuracy as an evaluation parameter in order to compare our model with other existing models. Since we used an almost balanced dataset, it does not affect the readings of accuracy very much. Other than accuracy we have used sensitivity, specificity \cite{r27} and Youden’s index (YI) \cite{r28} as evaluation metrics, to analyze our model performance. The sensitivity (true positive rate-TPR) of a classification model can be written as;

\begin{equation}
    TPR = \frac{TP}{TP+FN} \mbox{\ } \mbox{\ }
\end{equation}

where TP is true positive and FN is false negative, whereas specificity (true negative rate-TNR) can be written as;

\begin{equation}
{TNR = \frac{TN}{TN+FP}  \mbox{\ } \mbox{\ } }
\end{equation}

where TN stands for true negative. Thus, the specificity represents the proportion of negative samples that were correctly classified, and the sensitivity is the proportion of positive samples that were correctly classified. Generally, we can consider sensitivity and specificity as two kinds of accuracy, where former relates the actual positive samples and later relates actual negative samples. However, both sensitivity and specificity can be used for evaluating the classification
performance with imbalanced data.
The other parameter that we used to evaluate our model performance is youden’s index. It can be  used to assess the discriminative power of the test. The formula of youden’s index combines the sensitivity and the specificity and it can be written as follows;

\begin{equation}
    YI = TPR + TNR - 1 \mbox{\ } \mbox{\ }
\end{equation}

The YI metric is ranged from 0 to 1, where 0 represents a poor test and 1 represents a perfect test. Moreover, the accuracy of a classification model can be written as,

\begin{equation}
 {ACC = \frac{TP+TN}{TP+TN+FP+FN}} \mbox{\ } \mbox{\ }
\end{equation}

where FP is false positive and TN is true negative.

\section{Results and Evaluation}
\subsection{Results and Evaluation of Lung segmentation model}

\begin{figure}[]
\begin{subfigure}{.5\textwidth}
  \centering
  \includegraphics[width=.8\linewidth]{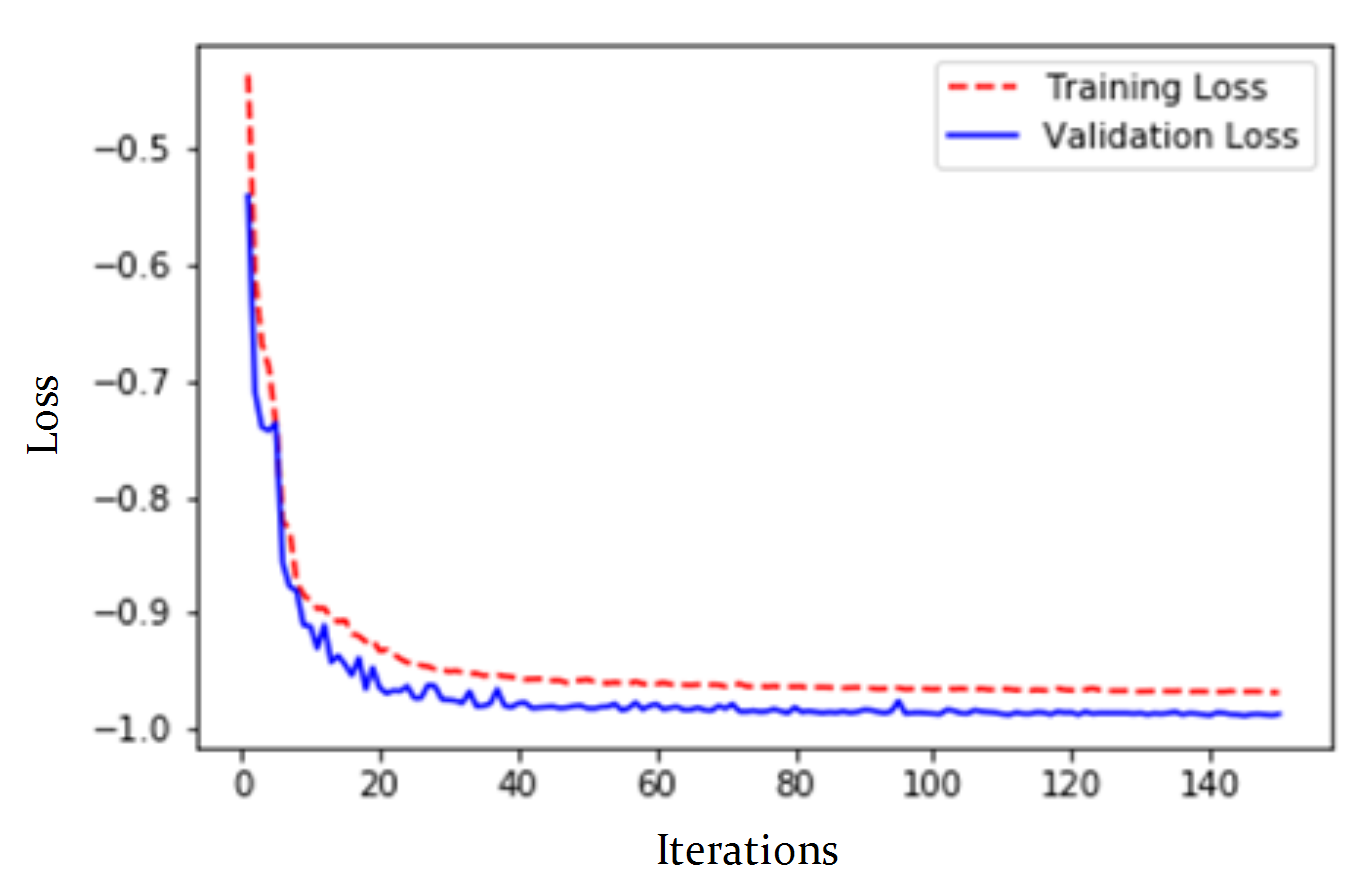}
  \caption{Training and validation losses}
  \label{fig3:sub-first}
\end{subfigure}
\begin{subfigure}{.5\textwidth}
  \centering
  \includegraphics[width=.8\linewidth]{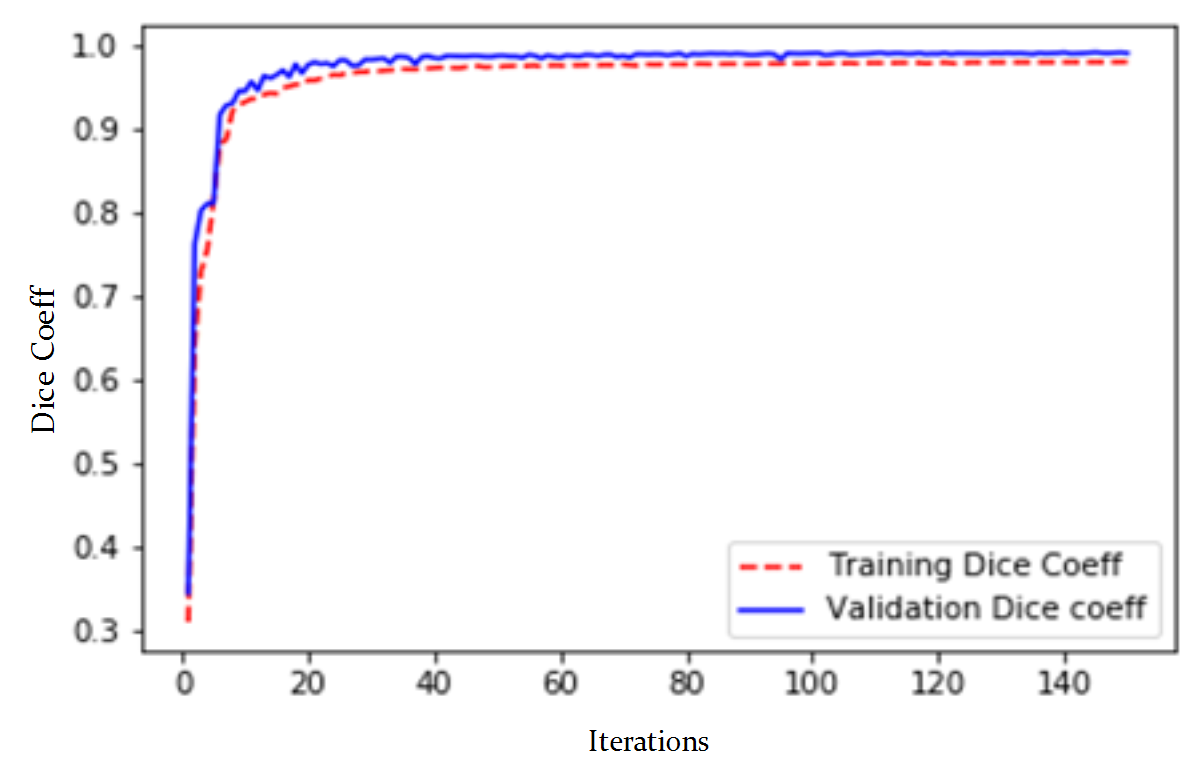} 
  \caption{Training and validation dice coefficient}
  \label{fig3:sub-second}
\end{subfigure}
\caption{Training curve of UNET model.}
\label{Training curve of UNET model.}
\end{figure}

%
%

As presented in Section 3.2, UNET is used to segment the lungs from CXR images. The training  curves, presented in Figure 5, indicate that the dataset size we used and the data augmentation techniques we used were sufficient to avoid overfitting. The mean DSC between gold standard segmentation and the UNET segmentation on the test set was 0.989. Several methods can be found in the literature to perform lung segmentation task such as rule-based method \cite{r29,r30}, deformable methods \cite{r31,r32}, hybrid models \cite{r33,r34} and deep learning models \cite{r35,r36}. The segmentation methods in \cite{r29}, \cite{r34}, \cite{r33}, \cite{r35} and \cite{r34} were able to achieve mean DSC of 0.88, 0.94, 0.946, 0.962 and 0.980 respectively. Our results show that lung segmentation model used in this research performs better than the other lung segmentation models discussed above.

\begin{figure}[]
  \centering
\begin{subfigure}{.5\textwidth}
  \center  
  \includegraphics[width=.85\linewidth]{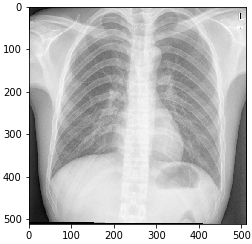}  
  \caption{input image.}
  \label{fig:sub-first2}
\end{subfigure}\hfill
\begin{subfigure}{.5\textwidth}
  \centering
    \includegraphics[width=.85\linewidth]{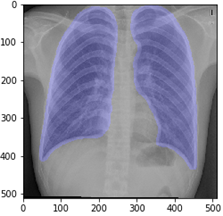}  
  \caption{UNET segmentation.}
  \label{fig:sub-second2}
\end{subfigure}
\begin{subfigure}{.5\textwidth}
  \centering
  \includegraphics[width=.85\linewidth]{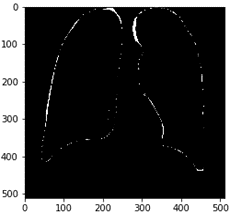}  
  \caption{Difference map .}
  \label{ fig:sub-three2}
\end{subfigure}
\caption{Comparison of original and predicted masks}
\label{fig:fig}
\end{figure}



%
%

The Figure 6 presents a comparison between the outcome of our model (6.b) and the original image (6.a) of the datasets.\hspace{1mm} 
The difference map (Figure 6.c) was generated from pixel wise subtracting of original image with UNET segmented image. Noted that, there were few discrepancies between two segmentations at the outer edges of the two lung fields, yet it does not impact the detection significantly as these areas are less likely to be infected by tuberculosis.

\subsection{Results and Evaluation of DCNN Model}

Figure 7 shows the training curve of the InceptionV3 DCNN.By observing the training curve it can be seen that the model has not overfit due to the strategies we used.


\begin{figure}[]
  \centering
  \includegraphics[width=1.0\linewidth]{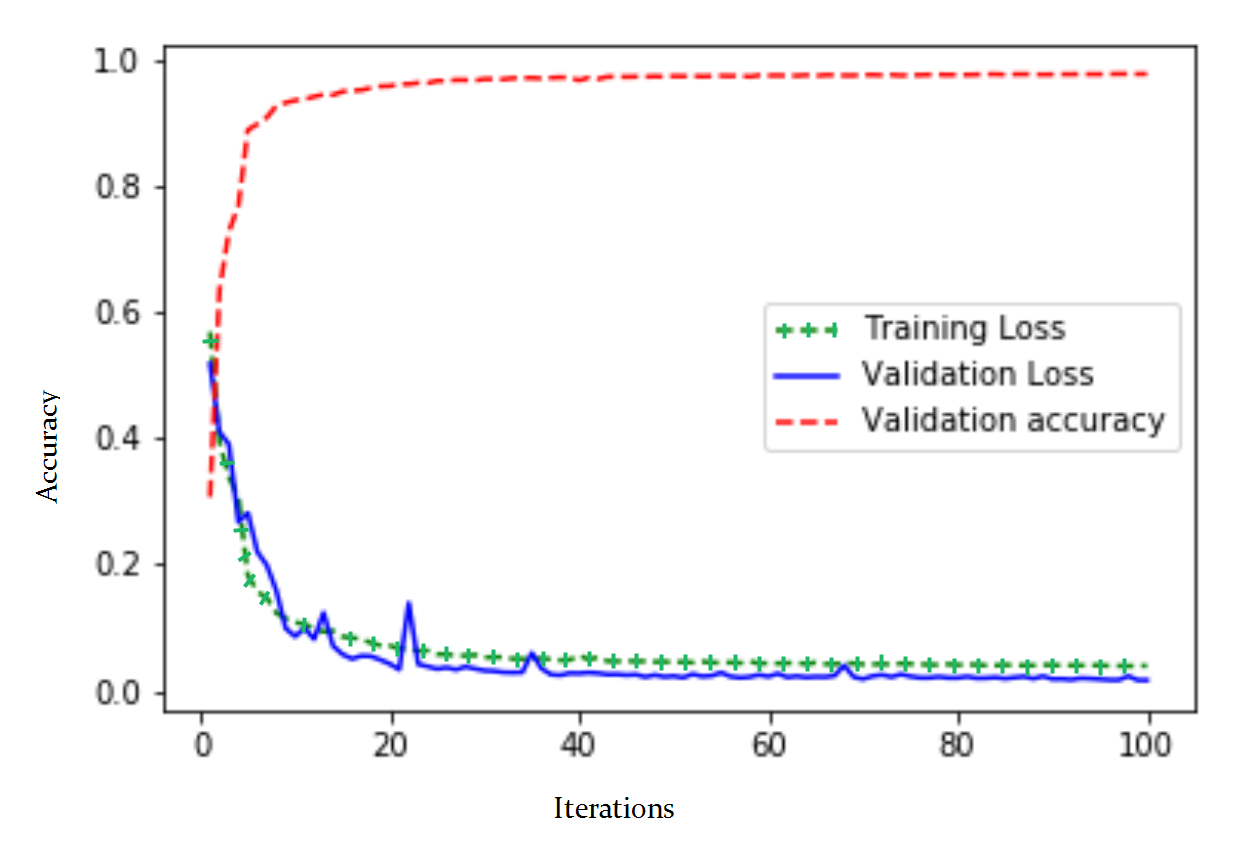}
  \caption{training curve of VGG16 DCNN}
  \label{fig5}
\end{figure}

%
%

We randomly chose 277 images (10\%) from real images of our complete dataset as testing dataset and it consists of 130 tuberculosis negative patients and 147 tuberculosis positive patients. To assess the overall performance of the classification youden’s index, sensitivity, specificity and accuracy were calculated using the results presented at Table 4 and Equation (5), (3) and (4) respectively. The calculated performance evaluation metrics were tabulated in Table 4. 

\setcounter{table}{2}
\begin{table}
\centering
\caption{training curve of VGG16 DCNN}
\begin{tabular}{|c|c|r|r|r|}
\hline
\multicolumn{2}{|l|}{}               & \multicolumn{1}{c|}{\begin{tabular}[c]{@{}c@{}}Diagnosis\\ +\end{tabular}} & \multicolumn{1}{c|}{\begin{tabular}[c]{@{}c@{}}Diagnosis\\ -\end{tabular}} & \multicolumn{1}{l|}{Total} \\ \hline
\multirow{3}{*}{VGG16}       & Test+ & 139                                                                         & 9                                                                          & 148                         \\ \cline{2-5} 
                             & Test- & 8                                                                          & 121                                                                         & 129                         \\ \cline{2-5} 
                             & Total & 147                                                                         & 130                                                                         & 277                         \\ \hline
\multirow{3}{*}{InceptionV3} & Test+ & 143                                                                         & 6                                                                          & 149                         \\ \cline{2-5} 
                             & Test- & 4                                                                          & 124                                                                         & 128                         \\ \cline{2-5} 
                             & Total & 147                                                                         & 130                                                                         & 277                         \\ \hline
\multirow{3}{*}{Ensemble}    & Test+ & 144                                                                         & 5                                                                          & 149                         \\ \cline{2-5} 
                             & Test- & 3                                                                          & 125                                                                         & 128                         \\ \cline{2-5} 
                             & Total & 147                                                                         & 130                                                                         & 277                        \\ \hline
\end{tabular}
\label{tab:addlabel}%
\end{table}



%


%

 %
%
%

\begin{table}
\centering
\caption{Performance of the Proposed Classification System }
\begin{tabular}{|l|l|l|l|l|}
\hline
            & \begin{tabular}[c]{@{}l@{}}Youden's\\  Index\end{tabular} & Sensitivity & Specificity & Accuracy \\ \hline
VGG16       & 0.875                                                     & 0.945       & 0.930       & 0.938    \\ \hline
InceptionV3 & 0.925                                                     & 0.972       & 0.953       & 0.963    \\ \hline
Ensemble    & 0.941                                                     & 0.979       & 0.962       & 0.971    \\ \hline
\end{tabular}
\end{table}

\vspace{3mm}

\begin{figure}[]
  \centering
	  \includegraphics[width=0.6\linewidth]{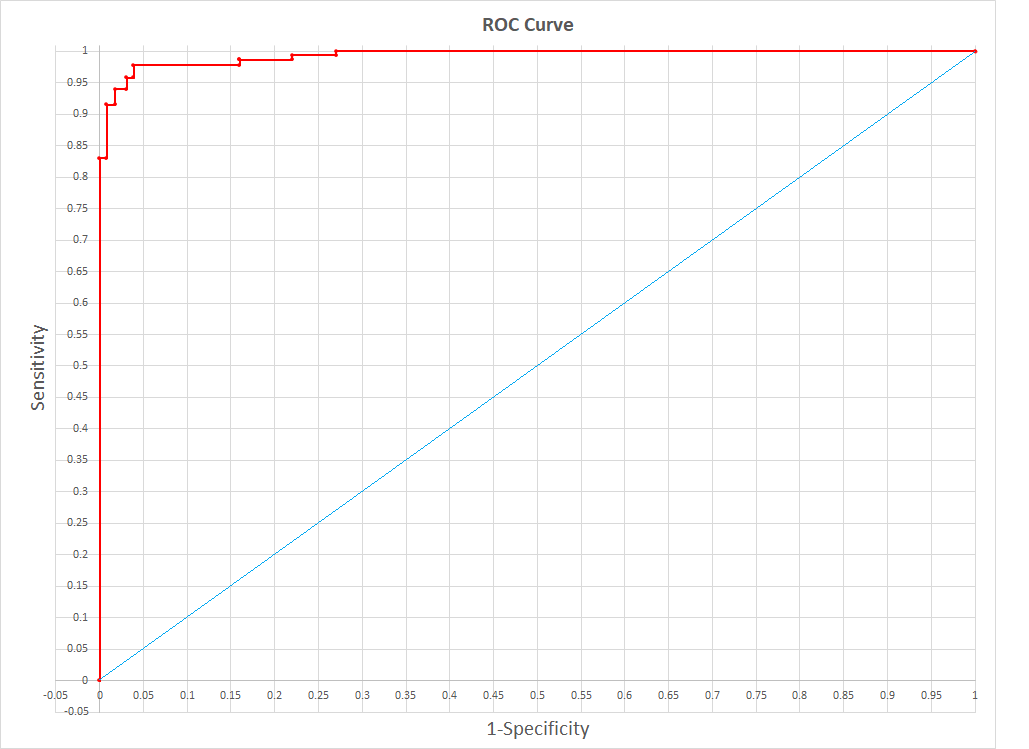} 
  \caption{ROC curve for ensemble model}
  \label{UNET}
\end{figure}

The ensembling was performed by taking different weighted averages of the probability scores generated by each classifiers. We put stronger weighting (60\%) on InceptionV3 model since it performs slightly better than VGG16 according to the results in Table 3. It can be seen that the ensemble performs better than VGG16 or InceptionV3 model alone and it was also able to achieve high sensitivity score as well, according to the Table 4. The Youden’s index is close to 1 in all the three models, and ensemble was able to achieve the best score of 0.941 among all. In \cite{r3,r4,r5} author’s claim that they were able to achieve 82.8\%, 82.35\%, 85.3\% of accuracy in their test sets respectively and we were able to achieve {97.1\% ( 95\% confident interval: 94.1\%, 99.4\%)} of accuracy with sensitivity of {97.9\% ( 95\% confident interval: 93.7\%, 100\%)} and specificity of {96.2\% ( 95\% confident interval: 90.9\%, 99.3\%)}  
on our test dataset which is a considerable amount of improvement. Even though statistics show good numbers still, it’s hard to figure out whether the model focuses on the rights features when it makes decisions. In order to confirm this we used CAM visualization which showcase the model’s vision.

\subsubsection{Class Activation Map (CAM)}

Deep learning networks are often considered as “black boxes” that does not offers any information about the features it has learned or about the segment of the input which provided significant details to the network to provide an accurate prediction. When a model fails, yet gives an accurate prediction at the validation stage, the overall output of the trained model at application and testing, often fail spectacularly without any warning or explanation. Hence, it is important to validate the trained model thoroughly, and if possible adopt a visualising technique to find the model tuned area/features.

CAM is one technique that can be used to get visual explanations of the predictions of convolutional neural networks. In other words CAM helps in the analysis of understanding as to what regions of an input image influence the convolutional Neural Network’s output prediction. The technique relies on the heat map representation which highlights pixels of the image that triggers the model to associate the image with a particular class. Hence, we make use of CAMs to analyse our model performance further.

 {According to the Figure 9, model was able to correctly identify the TB infected area of the CXR image in most instances accurately. The red square indicates the abnormal region of the CXR which was identified as TB infected by the radiologist and this matches perfectly with the blue area of the CAM visualisation, which indicate the region which makes a significant contribution to the final decision. Yet, it should be stressed that although CAM analysis was positive for some TB negative images, our model correctly classified the image as TB negative.}

\begin{figure}[ht] 
  \begin{subfigure}[b]{0.5\linewidth}
    \centering
    \includegraphics[width=0.75\linewidth]{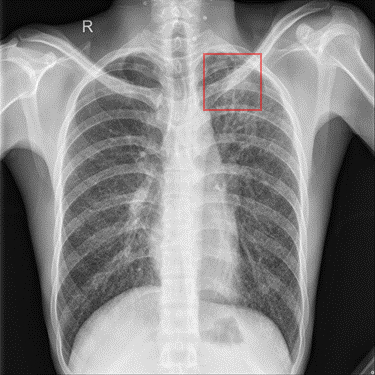} 
    \caption{TB positive X-ray} 
    \label{fig7:a} 
    \vspace{4ex}
  \end{subfigure}
  \begin{subfigure}[b]{0.5\linewidth}
    \centering
    \includegraphics[width=0.75\linewidth]{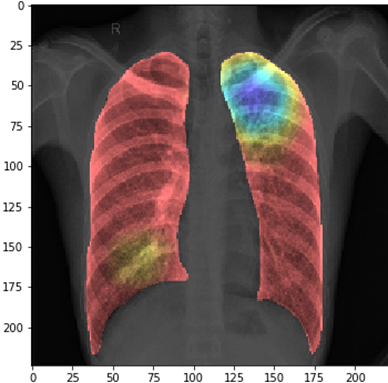} 
    \caption{CAM visualization of TB positive X-ray } 
    \label{fig7:b} 
    \vspace{4ex}
  \end{subfigure} 
  \begin{subfigure}[b]{0.5\linewidth}
    \centering
    \includegraphics[width=0.75\linewidth]{normal-chest-x-ray.jpg} 
    \caption{TB negative X-ray} 
    \label{fig7:c} 
  \end{subfigure}
  \begin{subfigure}[b]{0.5\linewidth}
    \centering
    \includegraphics[width=0.75\linewidth]{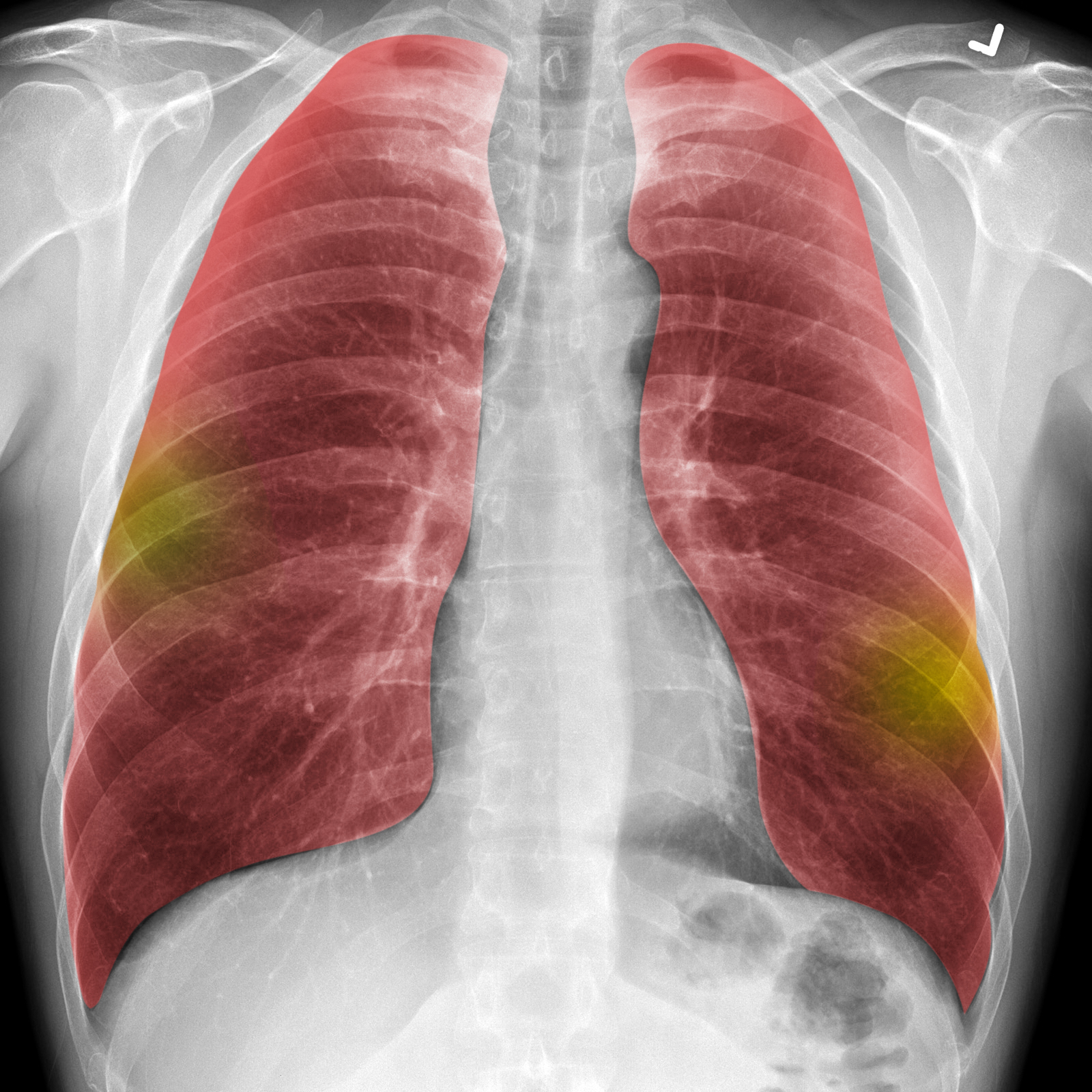} 
    \caption{CAM visualization of TB negative X-ray} 
    \label{fig7:d} 
  \end{subfigure} 
  
  \smallskip
  \centering{
\small { \textbf {Note:}  The red square in (a) shows the apical consolidation diagnose by the radiologist}}
  \caption{CAM visualization for TB positive and TB negative chest X-rays}
  \label{fig7} 
\end{figure}


%

\section{Discussion}
No ideal computer assisted tools are available so far to diagnose TB with high sensitivity and specificity. From the results it can be seen that the deep learning workflow we used in our model, mainly data preprocessing techniques, lung segmentation, hyper parameter optimization, image augmentation techniques and model ensembling have highly impacted on the model’s performance. {DC GAN based TB positive frontal x-ray image synthesis helps greatly to generalize the model and reduce overfitting. However, too much data augmentations such as random cropping, random brightness adjustments, random black patches, random Gaussian noise, cause poor accuracy, while grid distortion and elastic transformation improve the accuracy significantly. The VGG16 and InceptionV3 alone was able to achieve accuracy of 93.8\% and 96.3\% which is also a considerable improvement compared to the existing models in the literature \cite{r3, r4, r5}. Apart from these techniques several loss functions such as hinge loss\cite{r38}, focal loss\cite{r39} and several optimizers had been tested in our model development stage. But non helped to improve the model accuracy. The best performing optimizer for the two networks was observed as “Adam optimizer”. Since the last layer of the two networks is a softmax layer, binary cross entropy worked very well as the loss function. This is because the logarithmic term in the cross-entropy cancels out the plateau that is present in the soft-max function. However, ensembling helps to improve further the performance of our final model by removing uncorrelated errors of individual classifiers by using averaging.}
Finally, we used CAM visualization to showcase the infected area of the TB infected patients. In general, CAM allows researchers to visualize which pixels of the original image are responsible for predicting the corresponding class. Moreover, it helped us to get an intuitions about our model and how it localizes the main features in the image. However, there are limitations in this work, and it’s hard to compare model performance with human performance even if it scores high sensitivity and specificity since the model has not been tested in the practical field yet. The other drawback of this work is similar radiographic appearance in the CXR image such as lung cancers and bacterial pneumonia, may get detected as TB by the model. By integrating patient’s history and clinical findings such as coughing blood and having HIV Aids may help to overcome this issue up to certain level. Using data mining techniques along with deep learning and clinical data, could be used to overcome this limitation. Nevertheless, the main idea of this work is to provide a system that detects TB with high sensitivity and specificity only using CXR’s, and our performance analysis results in section 4.2 shows some promising outcome with  accuracy of 97.1\%,  sensitivity of 97.9\%, and specificity of 96.2\%.  Furthermore, the confidence interval (CI) values for ensemble model for Sensitivity 95\% CI - (93.1\% - 99.2\%), Specificity 95\% CI - (89.9\% - 95.7\%) and Accuracy 95\% CI - (90.4\% - 97.8\%).

\section{Conclusion}
The research presents an accurate model for automated TB detection system. {In order to achieve our performance goal, four different state of the art convolution neural networks architectures (DC GAN, UNET, VGG16, InceptionV3) along with several augmentation and preprocessing techniques were used. Among these techniques image synthesis, image preprocessing, lung segmentation, data augmentation, hyper parameter tuning and model ensembling made high impact on the final design. The lung segmentation task was carried out using UNET CNN and the tedious task of hyper parameter tuning was performed by using genetic algorithm. For the classification of chest X-rays showing radiological features of tuberculosis, we have used ensemble of two modified DCCNs and our detection accuracy in average is 0.971 while sensitivity is 0.979 and specificity is 0.962. Further, the model presented have a Youden’s index, which is a measure of overall predictive power of the diagnostic tool, of 0.941. However the detection accuracy may increase further by stacking more DCNN architectures such as Resnet, EfficientNet, Alexnet, and meta learning techniques could also be adopted to improve the performance of the model.} Finally we can conclude that the automated TB detection model presented in this work is reliable. Furthermore, according to the literature there does not exist an ideal computer assisted tool to diagnose TB with high sensitivity and specificity. Hence, the model presented could be quite helpful for the low incoming countries who are suffering from TB disease because the solution we have provided is inexpensive easily accessible and highly accurate, and can be used to screen large population instantly. 


%


 \section*{Compliance with ethical standards}
  \subsection*{Conflict of interest:}
The authors declare that they have no conflict of interest.

\subsection*{Ethical approval:}
For this type of study formal consent is not require . 

\subsection*{Informed consent:}
The dataset used in this article is a freely available. Both datasets were de-identified by the data providers and were exempted from IRB review at their respective institutions. At, NIH the dataset use and public release were exempted from IRB review by the NIH office of Human Resource Research Projection Programs (No. 5357).
Ref: Jaeger, Stefan, Sema Candemir, Sameer Antani, Yì-Xiáng J. Wáng, Pu-Xuan Lu,and George Thoma."Two public chest X-ray datasets for computer-aided screen-ing of pulmonary diseases." Quantitative imaging in medicine and surgery. 4.6(2014): 475.
\subsection*{Funding:}
This study was not funded by any grant company or person.

\end{document}